\documentclass[journal]{IEEEtran}
\usepackage{amssymb}
\usepackage{amsmath}
\usepackage{booktabs}
\usepackage{subfigure}
\usepackage{graphics}
\usepackage{graphicx}
\usepackage{hyperref}
\usepackage{}
\usepackage{amsmath} 

\ifCLASSINFOpdf
\else
\fi
\begin{document}
%
\title{Muscle Anatomy-aware Geometric Deep Learning for sEMG-based Gesture Decoding}
%
%
%


\author{Adyasha Dash*,
        Giulia Zappoli*,
        Laya Das, 
        Robert Riener,~\IEEEmembership{Senior Member,~IEEE}
\thanks{A. Dash, G. Zappoli and R. Riener are with the Sensory Motor Systems Lab, Department of Health Sciences and Technology, ETH Zurich, R. Riener is also with SCI Center, Medical Faculty, University of Zurich, and Distinguished Professor of AI and Robotics, Graduate School of Engineering, Tohoku University, Sendai, Corresponding author's e-mail: addash@ethz.ch).}
\thanks{L. Das was with the Department of Mechanical and Process Engineering, ETH Zurich}
\thanks{*These authors have equally contributed to the manuscript and share the first authorship.}}

%
%

\markboth{Journal of \LaTeX\ Class Files,~Vol.~14, No.~8, August~2015}%
{Shell \MakeLowercase{\textit{et al.}}: Bare Demo of IEEEtran.cls for IEEE Journals}
%



\maketitle

\begin{abstract}
Robust and accurate decoding of gesture from non-invasive surface electromyography (sEMG) is important for various applications including spatial computing, healthcare, and entertainment, and has been actively pursued by researchers and industry. Majority of sEMG-based gesture decoding algorithms employ deep neural networks that are designed for Euclidean data, and may not be suitable for analyzing multi-dimensional, non-stationary time-series with long-range dependencies such as sEMG. State-of-the-art sEMG-based decoding methods also demonstrate high variability across subjects and sessions, requiring re-calibration and adaptive fine-tuning to boost performance. To address these shortcomings, this work proposes a geometric deep learning model that learns on symmetric positive definite (SPD) manifolds and leverages unsupervised domain adaptation to desensitize the model to subjects and sessions. The model captures the features in time and across sensors with multiple convolutional kernels, projects the features onto SPD manifold, learns on manifolds and projects back to Euclidean space for classification. It uses a domain-specific batch normalization layer to address variability between sessions, alleviating the need for re-calibration or fine-tuning. Experiments with publicly available benchmark gesture decoding datasets (Ninapro DB6, FlexWear-HD) demonstrate the superior generalizability of the model compared to Euclidean and other SPD-based models in the inter-session scenario, with up to $\mathbf{8.83}$ and $\mathbf{4.63}$ points improvement in accuracy, respectively. Detailed analyses reveal that the model extracts muscle-specific information for different tasks and ablation studies highlight the importance of modules introduced in the work. The proposed method pushes the state-of-the-art in sEMG-based gesture recognition and opens new research avenues for manifold-based learning for muscle signals.  

\end{abstract}

\begin{IEEEkeywords}
Domain adaptation, gesture decoding, Riemannian manifold, generalizability.
\end{IEEEkeywords}

%
\IEEEpeerreviewmaketitle

\section{Introduction}




\IEEEPARstart{T}{he} human hand is one of the most versatile and dexterous parts of the human body: it performs numerous activities and gestures/grips achieved through a variety of motion strategies. With rapid development in the fields of wearable sensing, computing, and mechatronics, a plethora of new applications has emerged to restore, enhance and augment the power of the human hand. For example, intelligent prostheses allow the control of artificial limbs by decoding movement attempts from muscle physiology \cite{krasoulis2017improved}. Human-computer interaction (HCI) applications span from entertainment (e.g., virtual reality (VR)-based immersive gaming \cite{rautaray2011interaction}), to healthcare (e.g., touchless HCI \cite{ruppert2012touchless} and VR-based rehabilitation \cite{saurav2018design}) to spatial computing (e.g., real-time gesture-based interaction \cite{cao2024unveiling}).

These applications decode the hand movement and drive a visual, auditory, or propioceptive feedback through peripheral interaction \cite{saurav2018design}. Movement (or hand gesture) decoding is achieved by processing signals acquired through sensors such as surface electromyography (sEMG) electrodes, camera, etc.
SEMG-based sensing has been gaining popularity because of its non-invasive nature, low latency in recording biosignals and the possibility to record peripheral volitional control signals. These features allow for easy integration with the virtual environment, seamless interaction with the user and a better sense of immersiveness. In addition, availability of low-cost, safe, easy-to-use, off-the-shelf sEMG armbands make it a more lucrative mode of sensing. Thus, considerable research efforts have been dedicated to develop sEMG-based gesture decoding algorithms.

\subsection{Literature review}
Gesture decoding generally involves two steps - feature extraction and classification. Feature extraction has been achieved in the literature through time-domain \cite{zhang2022compound}, frequency-domain \cite{qi2019intelligent,hashi2024systematic}, or time-frequency \cite{hashi2024systematic} analysis. Classification has been achieved with linear discriminant analysis \cite{qi2019intelligent}, support vector machines \cite{hashi2024systematic}, etc. However, the sEMG signal, which captures motor unit action potentials (MUAPs), is a non-stationary time series corrupted with sensor noise and movement artifacts, leading to low signal-to-noise ratios (SNR) \cite{korovsec2000parametric}. Thus, the time-domain and frequency-domain techniques are ill-suited for this problem. Further, time-frequency analysis (wavelet)-based methods have limitations in capturing long-range dependencies such as slow drifts, fatigue trends, etc. \cite{al2011review}, and are sensitive to modeling choices such as order and type of wavelet \cite{al2011review}. Moreover, the suitability of the above hand-crafted features is greatly dependent on their ability to capture discriminative features from complex action-specific muscle activations and their design requires considerable heuristic knowledge.


Neural network-based methods offer an end-to-end approach for automated feature extraction and classification without the need for human engineering. Convolutional neural networks (CNNs) have been used to capture temporospatiofrequency features through hierarchical feature representation \cite{park2016movement}. The superior accuracy of CNNs over classical methods has been well-established in the literature \cite{park2016movement,oyemakinde2025novel}. Recently, Oyemakinde et al. proposed a novel attention-driven CNN to classify fine-finger gestures by fusing sEMG with force myography signals and reported state-of-the-art (SOTA)  performance \cite{oyemakinde2025novel} . Researchers have also combined CNN with long short-term memory (LSTM) units \cite{le2025cross,shin2025electromyography} to handle long-term dependencies in data. For example, \cite{le2025cross} developed a multi-stream CNN-BiLSTM architecture to classify complex hand movements by processing each channel separately with CNN and combining them through a bidirectional LSTM. \cite{shin2025electromyography} proposed a multi-branch model, with BiLSTM, CNN, and Bi-TCN (Temporal convolution network) branches, to combine bidirectional temporal relationships and time-varying patterns with past-and-future temporal context to improve the classification performance. Although these works have achieved SOTA performance, their efficiency has been reported to drastically decrease in cross-session scenarios \cite{le2025cross}, or has not been investigated in such settings \cite{oyemakinde2025novel,shin2025electromyography}.

\subsection{Research gaps} \label{sec:gaps}
Despite a range of methods spanning traditional techniques to neural network-based end-to-end learning, crucial research gaps still remain largely unaddressed, which are discussed below.

\subsubsection{Non-Euclidean structure of sEMG}
CNNs are primarily designed for computer vision tasks that assume the input data to be stationary, translation invariant, translation equivariant and stable with respect to local deformations \cite{ju2022tensor}, properties that can be verified for data with a Euclidean structure. The same applies for LSTM architectures. However, the sEMG signal, generated as a weighted summation of MUAPs, may not conform to this geometry \cite{gowda2024topology}. Thus, neural networks built for the Euclidean space can fall short of adequately capturing the relevant information in sEMG. Graph neural networks (GNNs) have been used as alternatives to Euclidean approaches \cite{massa2023explainable,lee2023stretchable}. For instance, \cite{massa2023explainable} used a GNN on high-density sEMG electrode arrays for decoding fine finger gestures and reported superior classification accuracy than conventional handcrafted algorithms. While GNN-based approaches \cite{massa2023explainable,lee2023stretchable} address the non-Euclidean structure of data, they have been evaluated on limited number of gestures \cite{lee2023stretchable}. Further, the design of the graph, specifically the edge weights, which highly affect the feature extraction and classification, is not straightforward for gesture recognition task.


\subsubsection{Inter-subject and inter-session variability}
An important challenge associated with processing sEMG is the high variability across subjects and sessions arising from extreme sensitivity to the positioning of electrodes, thickness of subcutaneous fat, spatial distribution of muscle fibers, distribution of muscle fiber conduction velocity, sweating, anatomical/physiological differences between subjects/sessions \cite{farina2004extraction,park2016movement}. Literature has shown that without proper domain adaptive approaches \cite{park2016movement} dedicated to handling this distribution shift, the decoding performance can decrease by up to $40\%$ \cite{du2017surface}. Although recalibration can address this, it is a repetitive and resource-intensive solution. Alternative approaches include adaptive learning \cite{park2016movement}, semi-supervised learning \cite{du2017semi} and domain adaptation \cite{yang2024emgbench}. \cite{park2016movement} proposed an adaptive learning method where the neural network is fine-tuned with a small dataset from an unseen user/session before the actual classification process. 
In \cite{yang2024emgbench}, the authors used correlation alignment for unsupervised domain adaptation, where the correlation matrix of input target data and the transformed source data is aligned in a latent space to minimize the distribution shift. They further fine-tuned with a small subset from unseen user data to achieve higher accuracy. These methods suffer from dependence on labeled data or pseudo-labeling schemes on the target domain to perform fine-tuning, which can be viewed as recalibration with a smaller dataset. Further, pseudo-labeling can spread error if the initial guessed lables are incorrect. In addition, these approaches employ Euclidean neural networks and do not consider the non-Euclidean structure of the data.

Thus, a gesture decoding framework that explicitly addresses the non-Euclidean structure of the data and offers generalizability across sessions/subjects without depending on (pseudo)labels is not available in the literature. This research gap reduces the efficiency and applicability to real-life applications. To address this gap, we propose  to use covariance-based features, such as common spatial patterns (CSP), based on eigen decomposition of covariance matrices, which have been proven to be superior for gesture classification \cite{riillo2014optimization, gowda2024topology}. 
While such manifold-based learning is better-suited to capture the non-Euclidean structure of the data, existing methods rely on hand-crafted features whose design can be inefficient for non-homogeneous sEMG sensor recordings and inadequate for modeling the domain shift across subjects/sessions. In this article, we draw inspiration from the success of learning on Riemannian manifolds and develop an end-to-end model coupled with unsupervised domain adaptation for gesture decoding.

\subsection{Contributions}
In this article, we propose a novel geometric deep neural network, called Temporal-Muscle-Kernel-Symmetric-Positive-Definite Network (TMKNet) for sEMG-based gesture classification. The contributions of this article are:
\begin{enumerate}
    \item An end-to-end deep neural network model that performs gesture classification in three steps: (1) feature extraction across time and sensors with multiple temporal and spatial kernels, (2) projection of features to the manifold of symmetric positive definite (SPD) matrices, learning on the SPD manifold and projecting back to the Euclidean space, and (3) classification of the hand movement.
    \item A multi-kernel spatial convolution layer that draws information from the underlying anatomy of gesture execution to extract relevant features for different types of movements.
    \item An unsupervised domain adaptation layer that uses domain-specific batch normalization (DSBN) to address domain shift across sessions/subjects.
    \item Demonstration of superior performance of TMKNet with two datasets (NinaproDB6 and FlexWear-HD) and generalizability of the model in the multi-session scenario.
\end{enumerate}

The remainder of the paper is organized as follows: Section \ref{sec:preliminaries} introduces the preliminary concepts. Section \ref{sec:method} describes the proposed method. The experimental setup is detailed in Section \ref{sec:experiment} and the results are provided in Section \ref{sec:result}. Section \ref{sec:discussion} provides a discussion and Section \ref{sec:conclusion} concluding remarks.

\section{Preliminaries} \label{sec:preliminaries}

\subsection{Riemannian geometry}

sEMG data exhibits non-Euclidean structure, and the notions of quantities like distance, mean and variance that are defined in the Euclidean space must be revised according to the structure of the data. We make use of the SPD manifold, which belongs to a class of manifolds called Riemannian manifolds and define these quantities.


A Riemannian manifold ($\mathcal{M}$) is a smooth manifold equipped with an inner product on the tangent space $\mathcal{T}_Z\mathcal{M}$ at each point $Z\in \mathcal{M}$. $\mathcal{T}_Z\mathcal{M}$ has a Euclidean structure with easy-to-compute distances that locally approximate Riemannian distances (on $\mathcal{M}$). 
Mathematically, the distance between two points $Z_1$ and $Z_2$ on $\mathcal{M}$
is defined as:
\begin{equation}
\delta(Z_1, Z_2) = \left\| \log\left(Z_1^{-\frac{1}{2}} Z_2 Z_1^{-\frac{1}{2}} \right) \right\|_F,
\end{equation}
where $||\cdot||_F$ is the Frobenius norm.


For a set of points $Z=\{Z_j\in\mathcal{M}\}$, the mean of $Z$, called Fréchet mean ($G_Z$) can be expressed as: 
\begin{equation}
G_Z=\arg\min_{G \in \mathcal{M}} \, \frac{1}{K} \sum_{j=1}^K \delta^2(G, Z_j).
\end{equation}
For $K=2$, $G_z$ has a closed-form solution:
\begin{equation}
G_Z(Z_1, Z_2; w) = Z_1 \#_w Z_2 = Z_1^{\frac{1}{2}}\left( Z_1^{-\frac{1}{2}} Z_2 Z_1^{-\frac{1}{2}} \right)^{w} Z_1^{\frac{1}{2}}.
\end{equation}
The variance of $Z$, called Fréchet variance ($V_Z^2$) can be expressed as:
\begin{equation}
V_Z^2 = \frac{1}{K} \sum_{j=1}^{K}\delta_{AIRM}^2(G,Z_j).      
\end{equation}
Here, $\#_w:$ denotes weighted geometric mean operator for the weight $w$.
These quantities are used in the forward and backward pass calculations in Riemannian neural networks. 

\subsection{Unsupervised domain adaptation}
Domain adaptation aims at building a model that is trained with data from one or more source domains, and can generalize to one or more target domains. Let $\textbf{S}=\{S_i\}_{i=1}^{N_s}$ represents the set of $N_s$ source domains and $\textbf{T}=\{T_j\}_{j=1}^{N_t}$ represents the set of $N_t$ target domains. Let us denote the input features and label for the $i^{th}$ source domain and $j^{th}$ target domain as $\{X_{S_i}, Y_{S_i}\}$ and $\{X_{T_j}, Y_{T_j}\}$, respectively. In this article, we adopt an unsupervised domain adaptation (UDA) approach, which aims to build a model that is trained with $\{X_{S_i}, Y_{S_i}\}_{i=1}^{N_s}$ and generalizes to all $X_{T_j}$ without using the label information $\{Y_{T_j}\}_{j=1}^{N_t}$. In contrast to (semi-)supervised domain adaptation, UDA does not rely on labels $\{Y_{T_j}\}_{j=1}^{N_t}$ or pseudo-labels, and thus, is cost-efficient, scalable to new domains and tasks, and has better real-world applicability, where labeled data for each new subject/session is unavailable.


\subsection{Riemannian Batch Normalization} \label{sec:rbn}
Batch normalization (BN) is a method that minimizes the covariate shift in the data and features by smoothing the distribution to speed up convergence. In the Euclidean space, BN involves centering and scaling operations realized through subtraction and division. However, on manifolds, these operations are achieved through parallel transport. For a vector $S$ on the tangent space of $Z_1$, the parallel transport operation  $\Gamma_{Z_1\rightarrow Z_2}(S)$ between $Z_1$, $Z_2$ $\in \mathcal{M}=\mathcal{S}^{++}$ (SPD manifold) defines a path from $Z_1$ to $Z_2$ through projection on tangent bundles (set of tangent spaces) while $S$ remains parallel to itself in the tangent spaces along the path as follows:
\begin{equation}
\forall S \in \mathcal{T}_{Z_1};\Gamma_{Z_1\to Z_2}(S) = (Z_2 Z_1^{-1})^{\frac{1}{2}} \, S \, (Z_2 Z_1^{-1})^{\frac{1}{2}} \in \mathcal{T}_{Z_2} 
\end{equation}

In the context of Riemannian batch normalization (RBN), the Fréchet mean of the $l^{th}$ batch $B_l$, denoted as $G_{B_l}$ can be approximated as proposed in \cite{brooks2019riemannian} by using the Karcher flow with only one iteration. Parallel transport can then be used to normalize with a bias parameter $G_\phi$ and variance parameter $V_\phi^2$ as follows \cite{brooks2019riemannian},]\cite{kobler2022spd}:
\begin{equation}
    \text{SPDBN}\left(Z_j;G_\phi,V_\phi^2,\epsilon\right)=\Gamma_{I\to G_\phi}\left(\Gamma_{G_{B_l} \to I}\bigl(Z_j\bigr)^{\frac{V_\phi}{V_{B_l}+\epsilon}}\right)  
\end{equation}

Batch-specific statistics during training can induce noise whose level depends on batch size, and is more pronounced for smaller batches. To combat this, SPD momentum batch normalization is used, where the algorithm uses two sets of statistics (Fréchet mean) separately for training and testing, which are updated with momentum parameters. In SPD momentum batch normalization (SPDMBN) \cite{kobler2022spd}, $\gamma_{source},\gamma_{target} \in [0,1]$ are clamped by exponential decay scheduler parameters for source and target domains.

\section{Methodology} \label{sec:method}
In this section, we propose a generalizable system-informed geometric deep neural network: TMKNet for sEMG-based gesture classification. TMKNet consists of three stages: (1) Multi-scale system-inspired feature extraction, (2) SPD Network and domain-specific batch-normalization, and (3) Classification, as shown in Fig. 1.

Let us denote an sEMG signal as $X\in\mathbb{R}^{c\times t}$ where $c$ and $t$ represent the number of channels (sensors) and number of samples in time, respectively. The proposed model takes as input a batch of examples $X_{batch}=\left[X_1, X_2, X_3, ..., X_{b}\right]\in\mathbb{R}^{b\times c\times t}$ and passes them through Euclidean and Riemannian layers to classify the movement. In the following, we present a detailed description of each layer of the proposed model.

\begin{figure*}
    \centering
    \includegraphics[width=\linewidth]{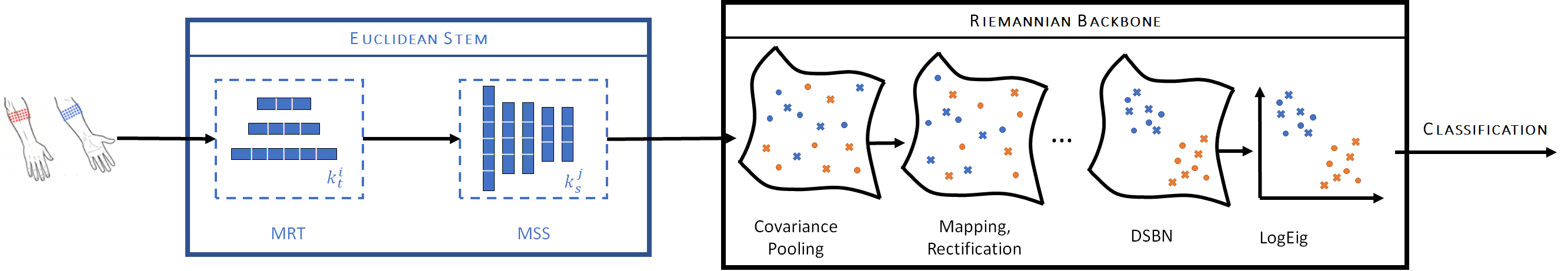}
    \caption{Proposed end-to-end Riemannian network for gesture classification from sEMG. The multi-channel data is fed to the Euclidean stem for multi-scale temporal (horizontal kernels) and spatial (vertical kernels) feature extraction with lengths $k_t^i$ and $k_s^j$, respectively. The features are transformed by the Riemannian backbone to SPD manifold, processed on manifolds and transformed back for final classification. Markers represent domains (subjects/sessions) and colours represent classes.}
    \label{fig:placeholder}
\end{figure*}

\subsection{Euclidean Stem}
The first module consists of multiple Euclidean layers that extract features at varying spatio-temporal resolutions from the sEMG signal. These layers are designed with inspiration from theory of muscle physiology, signal processing and human anatomy. 

\subsubsection{Multi-resolution temporal layer (MRT)}
As the sEMG signal is a non-stationary signal with varying frequency content, the time-frequency information at varying resolutions can provide important discriminative power for classifying the signal. The MRT layer is a composite layer consisting of 1-D convolution layers with different kernel sizes designed to extract this information at different scales. The sizes of temporal kernels are calculated from the sampling frequency ($Fs$) of the signal:
\begin{equation}
    k_t^i = (1, \alpha_iFs)
\end{equation}
where, $i$ represents the index of the kernel,  ${\alpha_i = {r_{data}}\cdot{r_{resolution,i}}}$ is the scaling factor for the $i^{th}$ kernel that includes dataset-specific and dataset-independent ratios, $r_{data}$ and $r_{resolution,i}$, respectively. Longer kernels can capture long-range dependencies and low frequency variations, while shorter kernels can extract high-frequency information.
By including multiple 1-D convolution layers with varying kernel sizes $k_t^i$, we aim to capture information at multiple frequency resolutions from the sEMG signal. The output of each kernel is passed through a \texttt{LeakyReLU} activation that introduces nonlinearity and is then downsampled using a \texttt{MaxPooling} layer to reduce the effect of noise and mitigate the curse of dimensionality. The outputs are stacked together along the feature dimension and passed through a Euclidean batch normalization layer to minimize covariate shift. The operations performed by the MRT layer can be expressed as:
\begin{eqnarray}
    z_t^i=&\texttt{MaxPool}\left(\texttt{LeakyReLU}\left(\texttt{Conv1D}\left(X, k_t^i\right)\right)\right)\\
    Z_t=&\texttt{BatchNorm}\left(\texttt{Stack}\left(\{z_t^i\}\right)\right)
\end{eqnarray}
The final output of the MRT layer is $Z_t\in\mathbb{R}^{b\times n_t\times c\times t_t}$, where $n_t$ represents the number of output channels. The last dimension is equal to the sum of the corresponding sizes of individual $z_t^i$. 

\subsubsection{Multi-scale Spatial Layer (MSS)}
At an anatomical level, different hand gestures are executed by the collective functioning of flexor and extensor muscles, that are manifested as different co-activation patterns in the sEMG signal. The multi-scale spatial layer is the second composite layer that extracts patterns from the output of the MRT layer, and is composed of the following kernels:
\begin{enumerate}
    \item Global muscle kernel, with size $k_s^1=(c, 1)$ that processes all the channels together.
    \item Flexor muscle kernel, with size $k_s^2=(c/2, 1)$ that processes channels corresponding to flexor muscles.
    \item Extensor muscle kernel, with size $k_s^3=(c/2, 1)$ that processes channels corresponding to extensor muscles.
    \item Proximal-Distal muscle kernel, with size $k_s^4=(c/2, 1)$ and a stride of $c/2$ that processes all proximal muscle sensors together, followed by all distal muscle sensors together.
    \item Dilated kernel, with size $k_s^5=(2, 1)$ and dilation factor of $c/2$.
\end{enumerate}
The outputs of all the layers are passed through \texttt{LeakyReLU} activation and then stacked together before  batch normalization. The operations performed by the MSS layer can be expressed as:
\begin{eqnarray}
    z_s^j&=\texttt{LeakyReLU}\left(\texttt{Conv1D}\left(X, k_s^j\right)\right)\\
    Z_s&=\texttt{BatchNorm}\left(\texttt{Stack}\left(z_s^1, z_s^2, ..., z_s^5\right)\right)
\end{eqnarray}
The final output of the MSS layer is $Z_s\in\mathbb{R}^{b\times n_s\times c_s\times t_t}$, where $n_s$ represents the number of output channels of the MSS layer. The last dimension is equal to the sum of the corresponding sizes of individual $z_s^j$.

\subsection{Riemannian Backbone}
The Euclidean stem extracts features from the sEMG signals that lie in the Euclidean space. The Riemannian backbone projects these Euclidean features onto a Riemannian manifold, specifically the SPD manifold, and then extracts features on manifolds while preserving the underlying geometric structure. In the following, we describe the different layers that constitute the Riemannian backbone.

\subsubsection{Covariance Pooling Layer}
To transform the Euclidean features, we calculate their covariance, that lies on the SPD manifold. 
The covariance matrix captures the second-order information in the extracted features and is a powerful tool to analyze multivariate sEMG data where functional connectivity between muscles groups carries vital information. This layer takes the flattened features $\tilde{Z}_s\in\mathbb{R}^{b\times n_s\times c_st_t}$ as input and computes the covariance matrix along the channel dimension. The covariance matrix is a positive definite matrix if the individual vectors are linearly independent. To ensure that the output of this layer always lies on the SPD manifold, a regularization is performed as follows:
\begin{eqnarray}
    \tilde{C}&=&\texttt{cov}(\tilde{Z}_s)\\
    C&=&\tilde{C} + \lambda I
\end{eqnarray}
The final output of the covariance pooling layer,  $C\in\mathcal{S}_{n_s}^{++}\subset\mathbb{R}^{b\times n_s\times n_s}$ lies on the SPD manifold $\mathcal{S}_{n_s}^{++}$.

\subsubsection{BiMap Layer}
The input covariance matrix can have high dimension due to a large number of features. The BiMap layer performs dimensionality reduction of covariance matrices while ensuring positive definiteness using a bilinear mapping as follows: 
\begin{equation}
    H_{bimap}=WCW^T
\end{equation}
The output of the BiMap layer, $H_{bimap}\in\mathcal{S}_{n_b}^{++}$ lies on the SPD manifold with $n_b<n_s$. In order to ensure that $H_{bimap}$ lies on an SPD manifold, the weight matrix $W$ is required to lie on a compact Stiefel manifold.

 
\subsubsection{ReEig Layer}
This layer is used to introduce non-linearity to the Riemannian module by eigenvalue rectification. It clamps the eigenvalues of the SPD matrices that are smaller than a threshold and then reconstructs the matrix. This is inspired by the rectified linear unit operation, common in Euclidean networks. This operation resembles nonlinear filtering on the input features, which can improve the expressive capacity of the module. The operations performed by the layer can be expressed as:
\begin{eqnarray}
    H_{bimap}&=&U_bS_bU_b^T\\
    H_{reeig}&=&U_b\max(\epsilon I_{n_b}, S_b)U_b^T
\end{eqnarray}
Here, $\epsilon$ is the rectification threshold and $I_{n_b}$ is the identity matrix of dimension $n_b$.

\subsubsection{Domain-specific Batch Normalization Layer}
This layer performs UDA on SPD manifold. This layer separates domain-specific information from the data in the form of BN statistics (mean and variance) and transforms the domain-specific data to domain-invariant data. It consists of multiple BN layers where each layer is devoted to one domain and captures domain-specific data statistics to normalize the data accordingly.

For a fixed batch size of $b$, batches are generally constructed with the same number of examples from different domains, so that the domains are uniformly represented. 
Since the batch size is fixed during training, 
an increase in the number of source domains leads to less number of samples per domain, and higher batch-specific noise. 
Thus, momentum DSBN is proposed, where two sets of running Fréchet means are computed for each domain to be used in training, and testing phase respectively. We construct this layer with multiple SPD batch normalization layers with momentum that have dedicated momentum parameters $\gamma_{source}$ and $\gamma_{target}$ for source and target domains, respectively. The operations performed by this layer can be expressed as:
\begin{align}
    H_{reeig}&=&\big[H_{reeig,S_1}, H_{reeig,S_2}, \ldots, H_{reeig,S_{n_s}},\nonumber 
    \\ & & H_{reeig,T_1}, H_{reeig,T_2}, \ldots, H_{reeig,T_{n_t}}\big] 
    \\ H_{bn,S_i}&=&\text{SPDMBN}\left(H_{reeig,S_i}, G_\phi,V_\phi^2,\epsilon,\gamma_{source}\right) \nonumber 
    \\ &  &\forall i\in\{1,2,3,\ldots,n_s\}
    \\ H_{bn,T_j}&=&\text{SPDMBN}\left(H_{reeig,T_j}, G_\phi,V_\phi^2,\epsilon,\gamma_{target}\right) \nonumber 
    \\ &  &\forall j\in\{1,2,3,\ldots,n_t\}
    \\ H_{dsbn}&=&\big[H_{bn,S_1}, H_{bn,S_2}, \ldots, H_{bn,S_{n_s}}, \nonumber 
    \\ & & H_{bn,T_1}, H_{bn,T_2}, \ldots, H_{bn,T_{n_t}}\big] 
\end{align}


    


\subsubsection{LogEig Layer}
This layer projects the SPD matrices onto the tangent space of $\mathcal{S}_{n_b}^{++}$, which is a Euclidean space.
This is achieved by replacing the eigenvalues of the matrix by their logarithms and reconstructing the matrix as follows:
\begin{eqnarray}
    H_{reeig}&=&U_rS_rU_r^T\\
    H_{logeig}&=&U_r\log(S_r)U_r^T
\end{eqnarray}
The final output of the module, $H_{logeig}\in\mathbb{R}^{b\times n_b\times n_b}$ is then processed by the classification head for gesture recognition.

\subsection{Classification head}
The final module consists of a flattening layer that transforms the matrix into a vector, followed by linear probing for classification, as follows:
\begin{eqnarray}
    l&=&\texttt{Linear}(\texttt{Flatten}(H_{logeig}))
\end{eqnarray}
The final output of this module, $l\in\mathbb{R}^{b\times n_c}$ represents the unnormalized logits for $n_c$ classes predicted by the model.


\section{Experimental set-up} \label{sec:experiment}
\subsection{Datasets and preprocessing}
We evaluate the proposed approach on two publicly available gesture recognition datasets: (1) NinaproDB6 \cite{palermo2017repeatability}, and FlexWear-HD \cite{yang2024emgbench}. Both the datasets contain data from multiple subjects and sessions with drift in distributions across sessions and subjects. These datasets have previously been cited as valuable resources for analyzing practical measures of out-of-distribution performance for sEMG datasets \cite{yang2024emgbench}.

\subsubsection{NinaproDB6} This dataset has $12$ repetitions (each repetition referred as a trial) of $7$ object grasping actions executed with $14$ real-life objects. The sEMG data was collected with $14$ Delsys Trigno sEMG electrodes with sampling frequency of $2000$ Hz. The sEMG sensors are placed in two rows in the forearm covering both flexor and extensor muscles. The dataset comprises 10 subject data with each subject recorded twice ($2$ sessions) per day over a period of $5$ days. 

We adopt the preprocessing approach presented in \cite{palermo2017repeatability}. At first, the sEMG data is synchronized. The data is then labeled and augmented with a $200$ ms rolling window with $100$ ms overlap; a Hampel filter is applied on each window to remove powerline interference at $50$ Hz. This process results in {$156408$} trials of $200$ ms each. The resulting trials are z-score normalized.

\subsubsection{FlexWear-HD} This dataset consists of $10$ hand gestures from $13$ subjects with $2$ sessions per subject in a single day. The sEMG data was collected using an array of $64$ hydrogel electrodes with gold-plated copper pads from the forearm covering both flexor and extensor muscles at $4000$ Hz. $8-10$ repetitions and $4-5$ repetitions per gesture were taken in the first and second session, respectively.
We adopt the preprocessing steps presented in \cite{yang2024emgbench}. Each trial passes through a $250$ ms sliding window with $125$ ms overlap for trial augmentation. Each resulting trial is then passed through a $50$ Hz Hampel filter for powerline noise cancellation and Z-score normalized. The total number of trials generated after augmentation is $31185$.

\subsection{Evaluation baseline, metrics and scenarios}
This article proposes the first geometric deep neural network designed for sEMG-based gesture decoding. Therefore, we compare its performance with a geometric network that was previously developed for motor imagery classification with EEG signals, called TSMNet \cite{kobler2022spd}. 

We evaluate and compare the performance of TMKNet using classification accuracy and F1 score to address the issue of class-imbalance in FlexWear-HD dataset. 
In order to compare the statistical relevance of the results, we used non-parametric Wilcoxon signed-rank test for F1 scores and accuracies of the TSMNet and TMKNet.

The efficacy of the TMKNet is evaluated in the inter-session setting with a leave-one-out approach, where training is performed on all sessions except one of a specific subject and the left-out session is used for testing. 

\subsection{Implementation details}
The code is implemented in PyTorch library and the source code is available at \href{https://github.com/gzappoli/UDA-sEMG}{\texttt{github.com/gzappoli/UDA-sEMG}}. The Euclidean stem consists of one MRT and one MSS layer with $r_{data}$ is set to $\frac{1}{5}$ for NinaproDB6 and $\frac{1}{4}$ for FlexWear-HD dataset, and $r_{resolution,i}$ is set to $\{\frac{1}{16}, \frac{1}{32}, \frac{1}{64}\}$, for $i=\{1,2,3\}$, respectively. For NinaproDB6 and FlexWear-HD respectively, the global muscle kernels length are $14$ and $64$, the flexor and extensor muscle kernel lengths are $7$ and $32$, the proximal-distal muscle kernel sizes are $7$ and $32$, the size of dilated kernel is $2$ with dilation factor of $7$ and $32$. In our experiments, $n_t=64$,  $n_s=40$ and $n_b=30$ for both datasets. The Riemannian backbone consists of one covariance pooling layer followed by a BiMap layer, a ReEig layer, a domain-specific batch normalization layer and a LogEig layer. Total number of parameters for Ninapro and FlexWear datasets are $84419$ and $346017$ respectively. 

The batch size is $50$ for both datasets and the number of epochs is $50$. The number of domains per batch is $5$ and $1$ for NinaproDB6 and FlexWear-HD respectively. The SPDDSMBN's momentum parameter is set to $\gamma_{source}=\gamma_{target}=0.1$ These hyperparameters are the same for all the subjects and sessions.
The RiemannianAdam optimizer from Geoopt toolbox \cite{geoopt2020kochurov} is used and the cross-entropy loss with learning rate $0.001$ and weight decay $0.0001$ is used for training. 


\section{Results} \label{sec:result}
In this section, we first present the performance of the proposed model and compare it with TSMNet \cite{kobler2022spd}. We then show a visual representation of the features extracted by the Euclidean stem and also highlight the role of the DSBN layer in performing unsupervised domain adaptation on Riemannian manifold. Finally, ablation studies demonstrate the importance of the different layers introduced in the model.


\subsection{Performance of TMKNet}
Table \ref{tab:performance} presents the classification accuracies and F1-scores of the proposed model, averaged across all participants for both datasets. TMKNet outperforms UDA-based methods in the literature that use Euclidean networks without target labels for fine-tuning ($0.6152\pm0.1159$, $0.6169\pm0.1191$, $0.6012\pm0.1282$ \cite{guo2024semg}) for the Ninapro-DB6 by at least $8.83$ points in accuracy. Other deep learning models that are not evaluated under true UDA, are excluded from the comparison. 

\begin{figure}
    \centering
    \includegraphics[width=\linewidth]{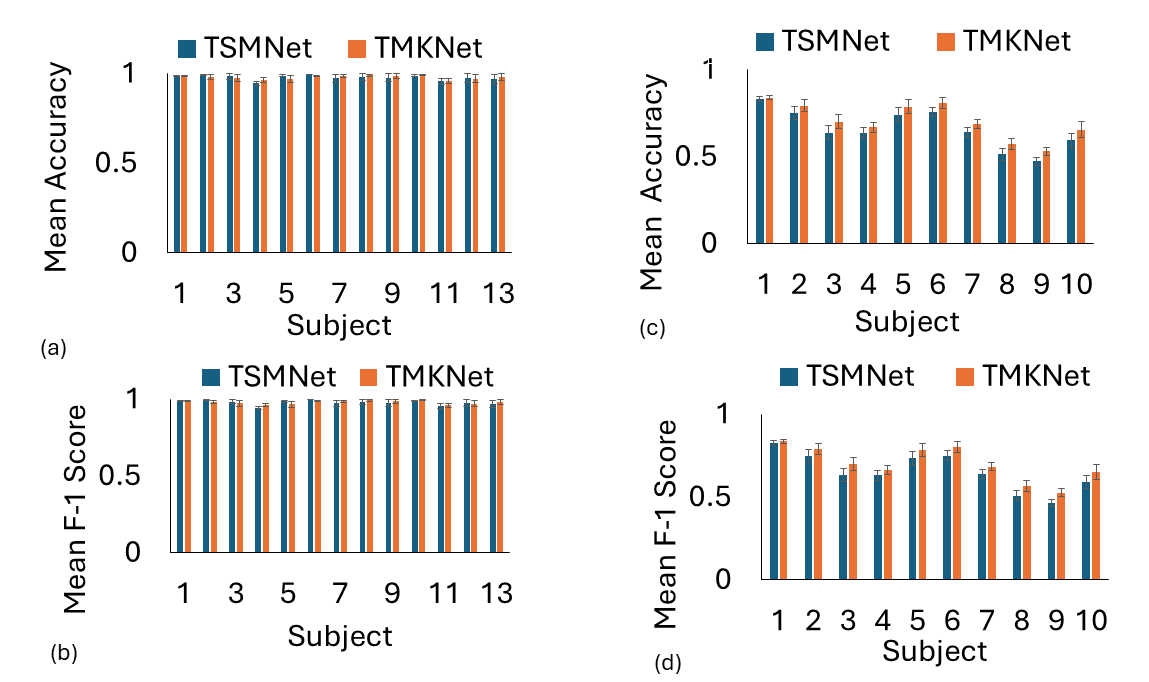}
    \caption{Subject-wise inter-session mean accuracy and F1 score with TMKNet and TSMNet (a)-(b) accuracy and F-1 score for FlexWear-HD, (c)-(d) accuracy and F-1 score for NinaproDB6.}
    \label{fig:acc}
\end{figure}

We also compare the performance of TMKNet with TSMNet for both datasets and report them in Table \ref{tab:performance}. We achieve an improvement of $4.63$ points in accuracy and $4.69$ points in F1-score for the NinaproDB6 dataset. The corresponding numbers are $0.15$ (accuracy) and $0.18$ (F1 score) for FlexWear-HD dataset. Furthermore, the standard deviation for accuracy is lower for the proposed model, implying higher robustness. A similar observation is found for the F1-score. The subject-wise average accuracies are shown in Fig. \ref{fig:acc}, which presents a more detailed view of the performance of the two models. It can be observed that, with the proposed model, there is an improvement in both metrics for all subjects in NinaproDB6 and majority of subjects in FlexWear.


\begin{table*}[h]
    \centering
    \caption{Inter-session classification accuracy and F1 score of TSMNet TMKNet for NinaproDB6 and FlexWear-HD datasets. \textbf{Bold} entries represent the best performance of the two models.}
    \begin{tabular}{c|c|c|c|c}
        \toprule
        Dataset & \multicolumn{2}{c|}{TSMNet} & \multicolumn{2}{c}{TMKNet}\\
        \cline{2-5}
         & Accuracy & F1 & Accuracy & F1 \\
         \midrule
        NinaproDB6 & $0.6623\pm0.1405$ & $0.6581\pm0.1408$ & $\mathbf{0.7086\pm0.1332}$ & $\mathbf{0.7050\pm0.1335}$\\
        FlexWear-HD & $0.9794\pm0.0211$ & $0.9787\pm0.0218$ & $\mathbf{0.9809\pm0.0168}$ & $\mathbf{0.9805\pm0.0170}$\\
        \bottomrule
    \end{tabular}
    \label{tab:performance}
\end{table*}

To investigate the statistical significance of the above results, we perform non-parametric Wilcoxon signed-rank test, considering the non-normal distribution of data. We observe that for NinaproDB6, both the metrics are statistically significantly better with TMKNet with $p<0.0001$. However, for FlexWear-HD dataset, the observed differences are not significant, potentially due to the lower number of sessions ($2$) compared to NinaproDB6 ($10$). 
Nevertheless, we observe improved accuracy and F1-scores for the majority of participants. 

\subsection{Interpretability}
\begin{figure}
    \centering
    \includegraphics[width=\linewidth]{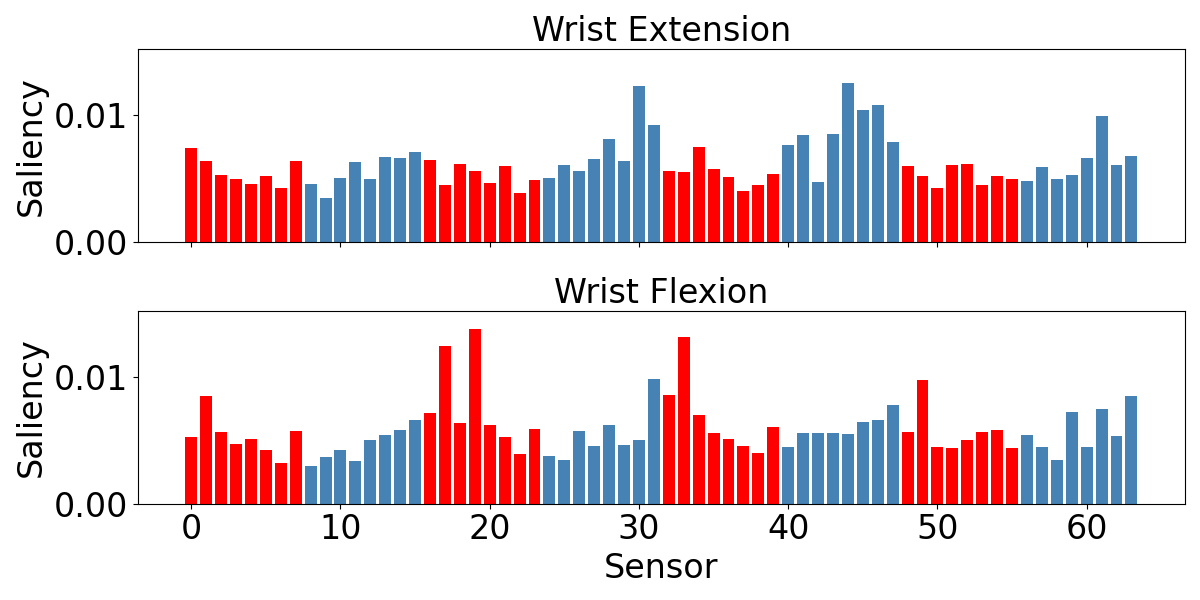}
    \caption{Sensor-wise maximum saliency for wrist extension and flexion tasks for a sample subject and session in FlexWear-HD dataset. Extensor and flexor sensors are shown in blue and red, respectively.}
    \label{fig:saliency}
\end{figure}

To interpret the features learnt by the model, we make use of saliency maps of the input channels. Fig. \ref{fig:saliency} plots the sensor-wise maximum saliency over time for two typical, complementary tasks, i.e., wrist extension and wrist flexion for subject 1, session 1 of FlexWear-HD. It can be observed from Fig. \ref{fig:saliency} that the saliency for extensor sensors are higher (in blue) for the extension task, while those for the flexion sensors (in orange) are higher for the flexion task.

A detailed analysis reveals that for the extension task, the top $10$ channels are found to be $30,44,31,46,47,25,45,61,34, 28$, all of which are located on the posterior side of the hand and capture sEMG from extensor muscles. Specifically, sensors $30, 31$ are placed (in middle) near extensor carpi radialis longus (ECRL) muscle which is important for functional wrist extension tasks. On the other hand, for flexion task, channels $19,39,1,20,31,17,49,61,47, 34$ are the top $10$ channels, of which, $7$ sensors are placed on the anterior side of the forearm, covering flexor muscles near flexor carpi radialis (FCR) and $3$ sensors are placed on the posterior side covering the extensor muscles. This is in line with the observation in the literature that for wrist flexion task, the flexor muscles are more active than the extensor muscles, with some extensors muscles activating as a synergist to stabilize the movement \cite{dash2019quantification}. Thus, the features extracted by the model are in agreement with the underlying mechanisms of hand movement.

\subsection{Effect of domain-specific batch normalization} 
\begin{figure}
    \centering
    \includegraphics[width=\linewidth]{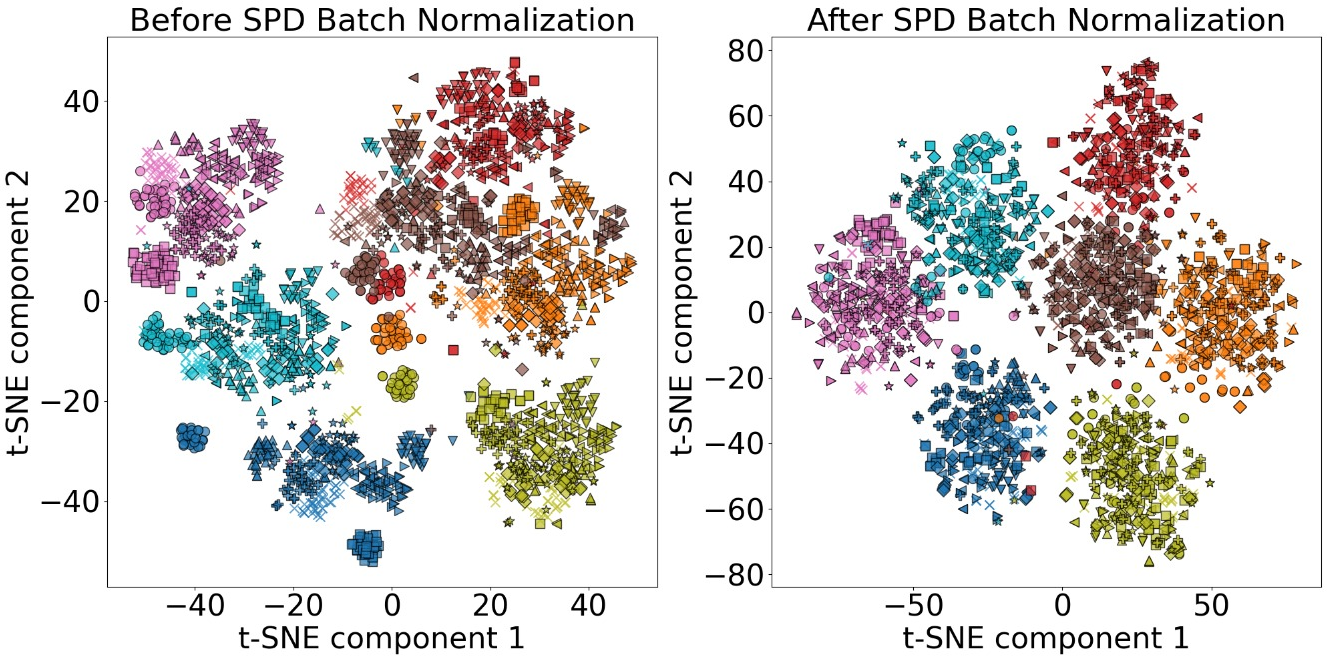}
    \caption{t-SNE of Riemannian features before and after domain-specific batch normalization for FlexWear-HD. Colour of points indicates class, marker indicates domain ($\star$: target domain, Others: source domains).}
    \label{fig:tsne}
\end{figure}

To highlight the importance of the DSBN layer in the TMKNet, we use t-distributed Stochastic Neighborhood Embedding (t-SNE) \cite{maaten2008visualizing} and generate $2D$ projections of the outputs of the intermediate layers in TMKNet. Fig. \ref{fig:tsne} compares the t-SNE before and after the DSBN layer for Ninapro. The plot comprises data from both source and target domains for subject $1$, with session $9$ as the target domain. It can be observed from Fig. \ref{fig:tsne} that prior to batch normalization, the samples within a class (e.g., blue data points) exhibit many patches, showing poor class-specific localization. Furthermore, for some classes, samples from different domains (markers) are segregated from others (e.g., cross marker for orange data points), showing poor domain-invariance of the features. The samples also overlap across classes (e.g., orange and red data points). After the DSBN layer, which performs normalization specific to each domain, the datapoints belonging to each class are better aggregated, forming localized clusters. Samples from different domains also exhibit better agreement with each other, demonstrating better domain-invariance. Finally, the overlap between classes is also reduced, allowing for accurate classification. 
This demonstrates the prowess of the DSBN layer in aggregating different domains (sessions) together, and thereby, leading to domain (session)-invariant learning and generalization across domains (sessions).


\subsection{Ablation Study}
Table \ref{tab:ablation} presents the results of ablation studies with respect to the layers introduced in the proposed model for subject $3$, session $7$ of NinaproDB6 and subject $7$ session $1$ of FlexWear-HD. The choice of the sessions (subjects) is based on the observation that their accuracies are close to the average accuracy in the respective datasets. 
For NinaproDB6, the overall accuracy decreases upon removal of any of the layers in the Euclidean stem. This decrease in performance is highest ($-0.2688$) when all kernels of the MSS layer are removed, followed by the case in which all kernels of MRT are removed ($-0.1535$). This shows the importance of MSS and MRT layers in the Euclidean stem. Further, it shows that the MSS layer designed in this article has higher contribution to the performance than the MRT layer. A kernel-wise analysis for the MSS shows that the largest performance loss is observed when either Flexor and Extensor muscle kernel or Proximal-Distal muscle kernel is removed. Thus, the muscle specific kernels, inspired from human anatomy are able to capture important discriminative information for gesture classification. A similar observation is found for F1-scores, where the metric decreases by removing any of the Euclidean layers. 

For FlexWear-HD, we observe that the removal of MRT or MSS layer reduces the performance, with highest drop for the MSS layer. Similarly, the removal of different kernels of the MSS layer reduces the performance, except for the dilated kernel. This observation points out that, while the proposed model produces a considerable overall performance improvement, the design of individual kernels and layers can benefit further by tuning them in a dataset-specific manner. 
\begin{table*}[]
    \centering
    \caption{Ablation study of TMKNet with respect to different layers and kernels in the Euclidean stem. The results are shown for a sample session from a subject and sample session in each dataset. \textbf{Bold} entries show the best performance.}
    \begin{tabular}{c|c|c|c|c}
        \toprule
        Model & \multicolumn{2}{c|}{Accuracy} & \multicolumn{2}{c}{F1 score}\\
        \cline{2-5}
         & NinaproDB6 & FlexWear-HD & NinaproDB6 & FlexWear-HD  \\
         \midrule
         TMKNet & $\mathbf{0.7249}$ & $0.9793$ & $\mathbf{0.7253}$ & $0.9791$\\
         w/o MRT & $0.5714$ & $0.9453$ & $0.5687$ & $0.9432$ \\
         w/o MSS & $0.4561$ & $0.6420$ &$0.4526$ & $0.6388$\\
         w/o Global muscle kernel & $0.7181$ & $0.9780$ & $0.7179$ & $0.9777$  \\
         w/o Flexor and Extensor muscle kernels& $0.6955$ & $0.9740$ & $0.6957$ & $0.9741$  \\
         w/o Proximal-Distal muscle kernel & $0.6955$ & $0.9727$ & $0.6957$ & $0.9725$ \\
         w/o Dilated kernel & $0.7105$ & $\mathbf{0.9813}$ & $0.7092$ & $\mathbf{0.9813}$ \\
        \bottomrule
    \end{tabular}
    \label{tab:ablation}
\end{table*}

\section{Discussion} \label{sec:discussion}

\begin{figure}
    \centering
     \includegraphics[width=0.8\linewidth]{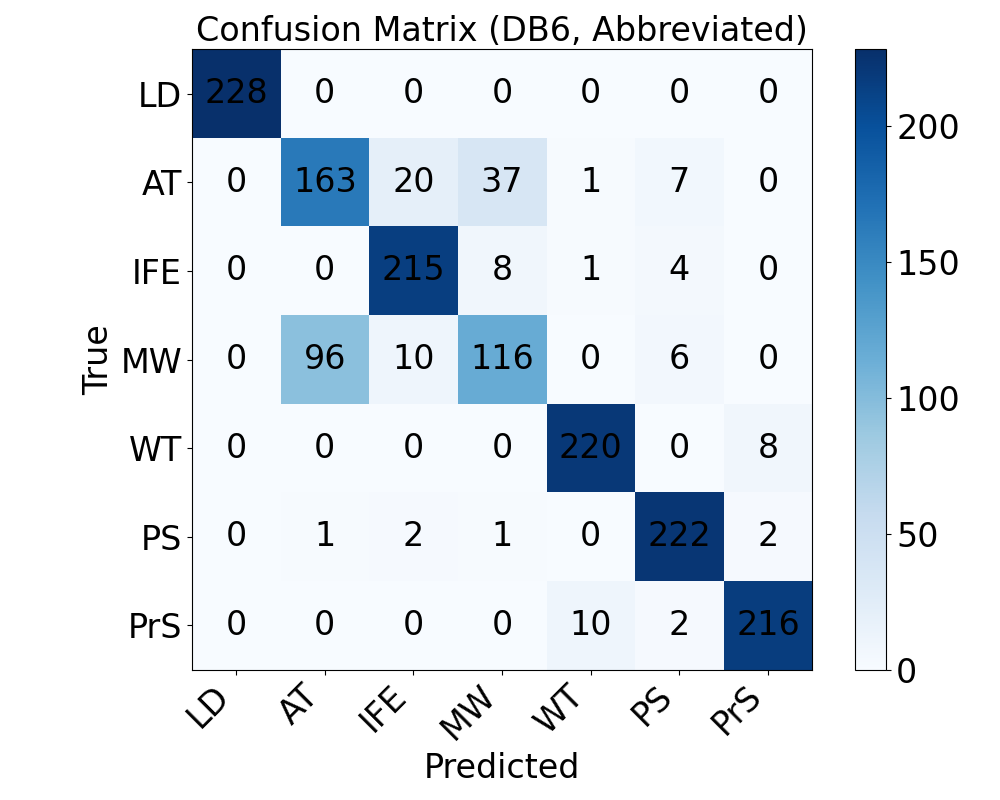}
    \caption{Confusion matrix for NinaproDB6 (subject 1, session 1). LD: Large Diameter, AT: Adducted Thumb, IFE: Index Finger Extension, MW: Medium Wrap, WT: Writing Tripod, PS: Power Sphere, PrS: Precision Sphere.}
    \label{fig:conf}
\end{figure}

\textit{Misclassifications by TMKNet:} We investigate the types of misclassifications performed by TMKNet and show the confusion matrix for a typical case (subject $1$, session $1$) of NinaproDB6 dataset in Fig. \ref{fig:conf}. It can be observed that the highest number of misclassifications are made for MW (medium wrap) movement, which are mostly misclassified as AT (adducted thumb). This is followed by misclassifications for AT, most of which are predicted as MW. It is interesting to note that MW and AT have similar gesture formation patterns, i.e., thumb-vs-all-finger closing and also have nearly similar thumb adduction \cite{palermo2017repeatability}. Thus, the mistakes made by the model are for the hand gestures that are closely related.


\textit{Inter-subject performance:} As discussed in Section \ref{sec:gaps}, the domain shift occurs across sessions and subjects. In this article, TSMNet was shown to provide the best performance for the inter-session scenario. We also trained TMKNet for the inter-subject scenario with both datasets, and found that the performance metrics were poorer compared to TSMNet. This can potentially be due to the choice of architecture that was optimized for inter-session performance, and further investigation is needed to correct this with a more detailed architecture search targeted at improving inter-subject performance. 
\section{Conclusion}\label{sec:conclusion}
In this article, we developed a geometric deep neural network,named TMKNet, for generalized sEMG-based gesture classification.

\textit{Key findings:} The proposed model, designed with anatomy-informed multi-kernel feature extraction coupled with learning on SPD manifolds outperforms Euclidean networks and Riemannian networks for gesture classification from sEMG data for the inter-session generalizability scenario. The model is able to extract task-specific contraction patterns from muscles that contract differently for different tasks. The domain-specific batch normalization plays a key role in clustering features from different domains, necessary for domain generalization, while preserving separability across classes, necessary for classification. Finally, the anatomy-informed multi-kernel MSS layer proposed in the article provides the highest gain in performance.

\textit{Outlook:} Euclidean neural networks have dominated the sEMG-based gesture decoding research. Geometric neural networks like TMKNet, operating on manifolds, can be robust for decoding hand gestures from sEMG. This work focused on inter-session generalizability with SPD layers coupled with UDA. This opens up new research avenues that can be pursued to improve the applicability in real-world settings. It was found that the same architecture does not work as well for the inter-subject scenario. Further modification might be needed in this direction. The datasets used for evaluating the model contain $64$ and $14$ sensors for different gestures, such as fine finger movements, grip and wrist flexion/extension. Further development with a wider variety of gestures is needed to evaluate the scalability of the model to real-life applications. While the model relies on UDA and does not need labels for the target domain, it still needs data for obtaining the batch statistics in the DSBN layer. This requirement can be further relaxed by leveraging test-time adaptation, where batch statistics can be updated in real time. The model struggles to accurately classify gestures that are very similar to each other, and additional sensors such as cameras can be used to improve the performance in a multi-modal framework.

\section*{Acknowledgments}
This work is funded by Swiss National Science Foundation Swiss Postdoctoral Fellowship under grant TMPFP3\_224549.

\ifCLASSOPTIONcaptionsoff
  \newpage
\fi



%
\bibliography{bibtex/bib/IEEEexample}

@misc{geoopt2020kochurov,
    title={Geoopt: Riemannian Optimization in PyTorch},
    author={Max Kochurov and Rasul Karimov and Serge Kozlukov},
    year={2020},
    eprint={2005.02819},
    archivePrefix={arXiv},
    primaryClass={cs.CG}
}

@article{maaten2008visualizing,
  title={Visualizing data using t-SNE},
  author={Maaten, Laurens van der and Hinton, Geoffrey},
  journal={Journal of machine learning research},
  volume={9},
  number={Nov},
  pages={2579--2605},
  year={2008}
}

@article{dash2019quantification,
  title={Quantification of grip strength with complexity analysis of surface electromyogram for hemiplegic post-stroke patients},
  author={Dash, Adyasha and Dutta, Anirban and Lahiri, Uttama},
  journal={NeuroRehabilitation},
  volume={45},
  number={1},
  pages={45--56},
  year={2019},
  publisher={SAGE Publications Sage UK: London, England}
}

@inproceedings{palermo2017repeatability,
  title={Repeatability of grasp recognition for robotic hand prosthesis control based on sEMG data},
  author={Palermo, Francesca and Cognolato, Matteo and Gijsberts, Arjan and M{\"u}ller, Henning and Caputo, Barbara and Atzori, Manfredo},
  booktitle={2017 International Conference on Rehabilitation Robotics (ICORR)},
  pages={1154--1159},
  year={2017},
  organization={IEEE}
}

@article{brooks2019riemannian,
  title={Riemannian batch normalization for SPD neural networks},
  author={Brooks, Daniel and Schwander, Olivier and Barbaresco, Fr{\'e}d{\'e}ric and Schneider, Jean-Yves and Cord, Matthieu},
  journal={Advances in Neural Information Processing Systems},
  volume={32},
  year={2019}
}

@article{guo2024semg,
  title={sEMG-based inter-session hand gesture recognition via domain adaptation with locality preserving and maximum margin},
  author={Guo, Yao and Liu, Jiayan and Wu, Yonglin and Jiang, Xinyu and Wang, Yalin and Meng, Long and Liu, Xiangyu and Shu, Feng and Dai, Chenyun and Chen, Wei},
  journal={International Journal of Neural Systems},
  volume={34},
  number={03},
  pages={2450010},
  year={2024},
  publisher={World Scientific}
}

@article{kobler2022spd,
  title={SPD domain-specific batch normalization to crack interpretable unsupervised domain adaptation in EEG},
  author={Kobler, Reinmar and Hirayama, Jun-ichiro and Zhao, Qibin and Kawanabe, Motoaki},
  journal={Advances in Neural Information Processing Systems},
  volume={35},
  pages={6219--6235},
  year={2022}
}

@inproceedings{du2017semi,
  title={Semi-Supervised Learning for Surface EMG-based Gesture Recognition.},
  author={Du, Yu and Wong, Yongkang and Jin, Wenguang and Wei, Wentao and Hu, Yu and Kankanhalli, Mohan S and Geng, Weidong},
  booktitle={IJCAI},
  pages={1624--1630},
  year={2017}
}

@article{riillo2014optimization,
  title={Optimization of EMG-based hand gesture recognition: Supervised vs. unsupervised data preprocessing on healthy subjects and transradial amputees},
  author={Riillo, Francesco and Quitadamo, Lucia Rita and Cavrini, Francesco and Gruppioni, Emanuele and Pinto, Carlo Alberto and Past{\`o}, N Cosimo and Sbernini, Laura and Albero, Lorenzo and Saggio, Giovanni},
  journal={Biomedical Signal Processing and Control},
  volume={14},
  pages={117--125},
  year={2014},
  publisher={Elsevier}
}

@article{yang2024emgbench,
  title={EMGBench: benchmarking out-of-distribution generalization and adaptation for electromyography},
  author={Yang, Jehan and Soh, Maxwell and Lieu, Vivianna and Weber, Douglas J and Erickson, Zackory},
  journal={Advances in Neural Information Processing Systems},
  volume={37},
  pages={50313--50342},
  year={2024}
}

@article{du2017surface,
  title={Surface EMG-based inter-session gesture recognition enhanced by deep domain adaptation},
  author={Du, Yu and Jin, Wenguang and Wei, Wentao and Hu, Yu and Geng, Weidong},
  journal={Sensors},
  volume={17},
  number={3},
  pages={458},
  year={2017},
  publisher={MDPI}
}

@article{farina2004extraction,
  title={The extraction of neural strategies from the surface EMG},
  author={Farina, Dario and Merletti, Roberto and Enoka, Roger M},
  journal={Journal of applied physiology},
  volume={96},
  number={4},
  pages={1486--1495},
  year={2004},
  publisher={American Physiological Society}
}

@article{lee2023stretchable,
  title={Stretchable array electromyography sensor with graph neural network for static and dynamic gestures recognition system},
  author={Lee, Hyeyun and Lee, Soyoung and Kim, Jaeseong and Jung, Heesoo and Yoon, Kyung Jae and Gandla, Srinivas and Park, Hogun and Kim, Sunkook},
  journal={npj Flexible Electronics},
  volume={7},
  number={1},
  pages={20},
  year={2023},
  publisher={Nature Publishing Group UK London}
}

@article{massa2023explainable,
  title={Explainable AI-powered graph neural networks for HD EMG-based gesture intention recognition},
  author={Massa, Silvia Maria and Riboni, Daniele and Nazarpour, Kianoush},
  journal={IEEE Transactions on Consumer Electronics},
  volume={70},
  number={1},
  pages={4499--4506},
  year={2023},
  publisher={IEEE}
}

@article{qi2019intelligent,
  title={Intelligent human-computer interaction based on surface EMG gesture recognition},
  author={Qi, Jinxian and Jiang, Guozhang and Li, Gongfa and Sun, Ying and Tao, Bo},
  journal={Ieee Access},
  volume={7},
  pages={61378--61387},
  year={2019},
  publisher={IEEE}
}

@article{zhang2022compound,
  title={Compound motion decoding based on sEMG consisting of gestures, wrist angles, and strength},
  author={Zhang, Xiaodong and Lu, Zhufeng and Fan, Chen and Wang, Yachun and Zhang, Teng and Li, Hanzhe and Tao, Qing},
  journal={Frontiers in Neurorobotics},
  volume={16},
  pages={979949},
  year={2022},
  publisher={Frontiers Media SA}
}

@article{ju2022tensor,
  title={Tensor-cspnet: A novel geometric deep learning framework for motor imagery classification},
  author={Ju, Ce and Guan, Cuntai},
  journal={IEEE Transactions on Neural Networks and Learning Systems},
  volume={34},
  number={12},
  pages={10955--10969},
  year={2022},
  publisher={IEEE}
}

@article{gowda2024topology,
  title={Topology of surface electromyogram signals: hand gesture decoding on riemannian manifolds},
  author={Gowda, Harshavardhana T and Miller, Lee M},
  journal={Journal of Neural Engineering},
  volume={21},
  number={3},
  pages={036047},
  year={2024},
  publisher={IOP Publishing}
}

@article{al2011review,
  title={A review of non-invasive techniques to detect and predict localised muscle fatigue},
  author={Al-Mulla, Mohamed R and Sepulveda, Francisco and Colley, Martin},
  journal={Sensors},
  volume={11},
  number={4},
  pages={3545--3594},
  year={2011},
  publisher={Molecular Diversity Preservation International (MDPI)}
}

@article{korovsec2000parametric,
  title={Parametric estimation of the continuous non-stationary spectrum and its dynamics in surface EMG studies},
  author={Koro{\v{s}}ec, Dean},
  journal={International journal of medical informatics},
  volume={58},
  pages={59--69},
  year={2000},
  publisher={Elsevier}
}

@article{oyemakinde2025novel,
  title={A novel sEMG-FMG combined sensor fusion approach based on an attention-driven CNN for dynamic hand gesture recognition},
  author={Oyemakinde, Tolulope Tofunmi and Kulwa, Frank and Peng, Xinhao and Liu, Yan and Cao, Jianglang and Deng, Xinping and Wang, Mengtao and Li, Guanglin and Samuel, Oluwarotimi Williams and Fang, Peng and others},
  journal={IEEE Transactions on Instrumentation and Measurement},
  year={2025},
  publisher={IEEE}
}

@inproceedings{park2016movement,
  title={Movement intention decoding based on deep learning for multiuser myoelectric interfaces},
  author={Park, Ki-Hee and Lee, Seong-Whan},
  booktitle={2016 4th international winter conference on brain-computer Interface (BCI)},
  pages={1--2},
  year={2016},
  organization={IEEE}
}

@article{shin2025electromyography,
  title={Electromyography-Based Gesture Recognition with Explainable AI (XAI): Hierarchical Feature Extraction for Enhanced Spatial-Temporal Dynamics},
  author={Shin, Jungpil and Miah, Abu Saleh Musa and Konnai, Sota and Hoshitaka, Shu and Kim, Pankoo},
  journal={IEEE Access},
  year={2025},
  publisher={IEEE}
}

@inproceedings{le2025cross,
  title={Cross-Day Myoelectric Gesture Recognition with Hybrid Multistream CNN-Bidirectional LSTM},
  author={Le, Hongquan and Spinks, Geoffrey M and in het Panhuis, Marc and Alici, Gursel},
  booktitle={2025 IEEE International Conference on Mechatronics (ICM)},
  pages={1--6},
  year={2025},
  organization={IEEE}
}

@article{hashi2024systematic,
  title={A systematic review of hand gesture recognition: An update from 2018 to 2024},
  author={Hashi, Abdirahman Osman and Hashim, Siti Zaiton Mohd and Asamah, Azurah Bte},
  journal={IEEE Access},
  volume={12},
  pages={143599--143626},
  year={2024},
  publisher={IEEE}
}

@article{krasoulis2017improved,
  title={Improved prosthetic hand control with concurrent use of myoelectric and inertial measurements},
  author={Krasoulis, Agamemnon and Kyranou, Iris and Erden, Mustapha Suphi and Nazarpour, Kianoush and Vijayakumar, Sethu},
  journal={Journal of neuroengineering and rehabilitation},
  volume={14},
  number={1},
  pages={71},
  year={2017},
  publisher={Springer}
}

@inproceedings{rautaray2011interaction,
  title={Interaction with virtual game through hand gesture recognition},
  author={Rautaray, Siddharth S and Agrawal, Anupam},
  booktitle={2011 International Conference on Multimedia, Signal Processing and Communication Technologies},
  pages={244--247},
  year={2011},
  organization={IEEE}
}

@article{ruppert2012touchless,
  title={Touchless gesture user interface for interactive image visualization in urological surgery},
  author={Ruppert, Guilherme Cesar Soares and Reis, Leonardo Oliveira and Amorim, Paulo Henrique Junqueira and de Moraes, Thiago Franco and da Silva, Jorge Vicente Lopes},
  journal={World journal of urology},
  volume={30},
  number={5},
  pages={687--691},
  year={2012},
  publisher={Springer}
}

@article{cao2024unveiling,
  title={Unveiling the Era of Spatial Computing},
  author={Cao, Hanzhong},
  journal={arXiv preprint arXiv:2405.06895},
  year={2024}
}

@inproceedings{saurav2018design,
  title={Design of a VR-based upper limb gross motor and fine motor task platform for post-stroke survivors},
  author={Saurav, Kumar and Dash, Adyasha and Solanki, Dhaval and Lahiri, Uttama},
  booktitle={2018 IEEE/ACIS 17th International Conference on Computer and Information Science (ICIS)},
  pages={252--257},
  year={2018},
  organization={IEEE}
}
\bibliographystyle{ieeetr}
\end{document}